\documentclass{svproc}
\usepackage{url}

\usepackage{amsmath}
\usepackage{amssymb}
\usepackage{graphicx}
\usepackage{csquotes}
\usepackage{multirow}
\usepackage{tabularx}
\usepackage{xcolor} 
\usepackage[colorlinks]{hyperref} 

\newcommand{\Mod}[1]{\ \mathrm{mod}\ #1}

\usepackage{tikz}

\newcommand{\sqdiamond}[1][fill=black]{\tikz [x=1.2ex,y=1.85ex,line width=.1ex,line join=round, yshift=-0.285ex] \draw  [#1]  (0,.5) -- (.5,1) -- (1,.5) -- (.5,0) -- (0,.5) -- cycle;}%
\newcommand{\MyDiamond}[1][fill=gray]{\mathop{\raisebox{-0.275ex}{$\sqdiamond[#1]$}}}

\newcommand{\tikzcircle}[2][red,fill=red]{\tikz[baseline=-0.5ex]\draw[#1,radius=#2] (0,0) circle ;}%

\newcommand*{\greysquare}{\textcolor{gray}{\blacksquare}}

\begin{document}
\mainmatter              
\title{On the Role of Incentives in Evolutionary Approaches to Organizational Design}
\titlerunning{Incentives in Evolutionary Organizational Design}  
%
\author{Stephan Leitner}
\authorrunning{Stephan Leitner} 
%
\tocauthor{Stephan Leitner}
\institute{Department of Management Control and Strategic Management,\\ University of Klagenfurt, Klagenfurt, Austria,\\
\email{stephan.leitner@aau.at}}

\maketitle              

\begin{abstract}
This paper introduces a model of a stylized organization that is comprised of several departments that autonomously allocate tasks. To do so, the departments either take short-sighted decisions that immediately maximize their utility or take long-sighted decisions that aim at minimizing the interdependencies between tasks. The organization guides the departments' behavior by either an individualistic, a balanced, or an altruistic linear incentive scheme. Even if tasks are perfectly decomposable, altruistic incentive schemes are preferred over individualistic incentive schemes since they substantially increase the organization's performance. Interestingly, if altruistic incentive schemes are effective, short-sighted decisions appear favorable since they do not only increase performance in the short run but also result in significantly higher performances in the long run.

\keywords{Time preferences, Guided self-organization, Agent-based simulation, Bayesian learning, $N\!K$-framework, Vickrey auction}
\end{abstract}
\section{Introduction}

Organizational design choices, such as the division of tasks into sub-tasks, have a significant impact on the functioning and the performance of organizations \cite{burton2018,doty1993}. The main challenge of organizational design arises from the correlation between the tasks in an organization's task space, usually referred to as interdependencies \cite{simon2019}. Picture a stylized organization in which several agents jointly perform a global task divided into sub-tasks, and the sub-tasks are assigned to organizational units. The solutions to the sub-tasks are, later, integrated around the completion of the global task. In the case of interdependencies between sub-tasks, the efficiency of the individual behavior of an organizational unit depends crucially on the other units' behaviors \cite{galbraith1974}, which is why some controls, such as incentive systems, are required to coordinate behaviors. Thus, fundamental choices in organizational design are concerned with dividing the global task into sub-tasks and with designing mechanisms that permit coordinated actions across sub-tasks, usually referred to as structure and coordination, respectively \cite{burton2018,rius2020}. 

There are two main world-views of organizational design. First, classic approaches follow the premise of the rational actor and postulate that organizational design is the result of \textit{deliberate} decisions \cite{tsoukas1993}, whereby managers are responsible for designing feasible organizational structures. Design choices might follow information processing or contingency perspectives \cite{lawrence1967,galbraith1974}. \footnote{For reviews see, for example, Burton and Obel \cite{burton2018}, Joseph et al. \cite{joseph2018}, and Visscher and Fisscher \cite{visscher2012}.} 
This is in line with Simon \cite{simon2019} and Romme \cite{romme2003}, who argue that organizations are artifacts that are conceptually developed by a designer \textit{before} implementation. Organizational design choices are concerned with aligning, amongst others, structures, processes, and practices, so that an organization can meet its objectives \cite{tsoukas1993}. Naturally, classical approaches have a strong focus on macro-level organizational forms \cite{joseph2018}.

Second, evolutionary approaches consider that organizational design is an emergent property. Following Tsoukas \cite{tsoukas1993}, in evolutionary approaches to organizational design, decisions are no longer the outcome of rational actors who make deliberate decisions. Rather, a shift from the macro-level to the micro-structures can be observed; the latter is often concerned with the mechanisms that drive the emergence of organizational design \cite{joseph2018}. In particular, evolutionary approaches consider non-rational actors who act almost randomly, while their behaviors are guided by regulative processes concerned with \enquote*{survival} in an evolutionary sense. These processes are often referred to as plastic control \cite{tsoukas1993,popper1978} or guided self-organization \cite{prokopenko2009}. 
Following evolutionary approaches, organizational design is a collective learning process in which the whole organization actively participates. Unlike the classical approach, evolutionary approaches integrate the phases of design, and implementation \cite{visscher2012}, and organizations are conceptualized as complex adaptive systems  (CAS) \cite{ethiraj2004} that are characterized by (i) multiple autonomous, heterogeneous, and interacting agents with diverse behaviors, (ii) individual and system memory that affects learning, (iii) and adaptation and self-organization mechanisms that drive the systems' evolution \cite{Leitner2015,richardson2001}. 

Previous research tends to recommend that an organization's formal structure should \textit{mirror} the technical dependencies of the task that the organization faces \cite{sanchez1996}, to assure that the organization performs efficiently. This recommendation is referred to as {mirroring hypothesis} and has its roots in Simon's work on decomposable complex systems \cite{simon1991}. One crucial assumption of the mirroring hypothesis is that the technical structure of the task (i.e., the structure of interdependencies) is accessible for all individuals. Previous studies often include this assumption and regard information about interdependencies to be exogenously given and, more importantly, publicly available. However, in reality, the nature of the work is mostly unclear and unknown \cite{raveendran2020}. 
The hypothesis has been criticized, based on empirical evidence, since technical dependencies might co-evolve with organizational structures. At the same time, there is also empirical evidence that supports it \cite{querbes2018,baldwin2014}. In the context of evolutionary approaches to organizational design, the necessity to analyze micro-level self-organization processes becomes particularly evident since individual time-preferences \cite{rajan2009}, incentive systems, and the departments' self-organization strategy might interact. If departments follow the mirroring hypothesis and \enquote*{evolutionary optimize} for task interdependencies, it might take some time until the design choices pay off in terms of utility, and utility is also affected by the incentive system active in the organization.
On the contrary, there might be other micro-level mechanisms that deviate from the logic of the mirroring hypothesis. These mechanisms, on the one hand, might lead to immediate increases in individual utility. However, on the other hand, they might be detrimental to organizational performance. 
 
Evolutionary approaches to organizational design and the mirroring hypothesis still lack a precise theoretical framework that could substantially stimulate further research \cite{querbes2018}.
This work is a first step towards the development of this framework.  The paper proposes a model of evolutionary organizational design and analyzes the organizational performance achieved when different micro-level processes are active. 
The remainder of the paper is organized as follows: The model is introduced in Sec. \ref{sec:model}. The results are presented and discussed in Sec. \ref{sec:results}, and Sec. \ref{sec:conclusion} concludes the paper. 

%
\section{Model}
\label{sec:model}
\begin{figure}
\vspace{-0.25cm}
\includegraphics[width=\textwidth]{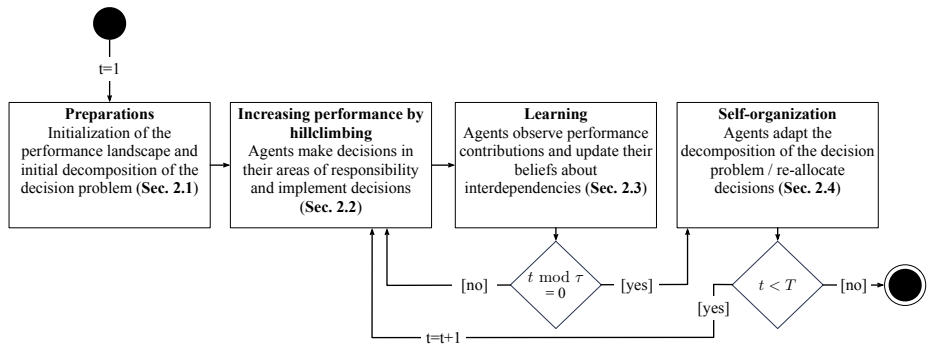}
    \caption{Flowchart}
    \label{fig:flowchart}
    \vspace{-0.25cm}
\end{figure}

The model of a stylized organization builds on the $N\!K$-framework \cite{levinthal1997,wall2020,blanco2020}. In particular, the model considers an organization that consists of $M\in \mathbb{N}$ agents that represent organizational units. Together, these agents face an $N$-dimensional decision problem with $K$ interdependencies among decisions, where $N\in \mathbb{N}$ and $K\in \mathbb{N}_0$. The interdependencies shape the decision problem's complexity. Agents cannot solve the decision problem alone but collaborate. To do so, they decompose the decision problem into sub-problems (Sec. \ref{sec:task-environment}) and employ a hillclimbing algorithm to increase their performance (Sec. \ref{sec:decisions}). Agents know that there might be interdependencies among decisions. However, they are not aware of the structure. Still, they are endowed with the capability to learn about the structure of interdependencies over time (Sec. \ref{sec:learning}), and they can adapt the task allocation following an auction-based mechanism (Sec. \ref{sec:auction}). For $t=\{1,\dots,T\}\subset \mathbb{N}$ periods it is observed, how learning and self-organization affect the performance. The sequence of events during a simulation run is illustrated in Fig. \ref{fig:flowchart}.

\subsection{Task Environment and Decomposition}
\label{sec:task-environment}
The decision problem faced by the agents consists of $N$ binary decisions and is formalized by 
\begin{equation}
\mathbf{d}=\left[d_1, d_2, \dots, d_N \right]~,
\end{equation}
where $d_n\in \{0,1\}$ and $n=\{1,\dots,N \} \subset \mathbb{N}$. Every decision $d_n$ contributes $f(d_n) \sim U(0,1)$ to the organization's performance. Due to interdependencies among decision, the performance contribution $f(d_n)$ might not only be affected by decision $d_n$ but also by $K$ other decisions. The corresponding contribution function for decision $d_n$ is formalized by 
\begin{equation}
    f\left(d_n\right)=f\left(d_n, d_{i_1}, \dots, d_{i_K} \right)~,
\end{equation}
where $\{i_1, \dots, i_K\} \subseteq \{1, \dots, n-1, n+1,\dots,N\}$ and $0\leq K \leq N-1$. The organizations' performance is computed according to Eq. \ref{eq:performance}, where $|\mathbf{d}|$ indicates the length of the vector $\mathbf{d}$:
\begin{equation}
    \label{eq:performance}
    P(\mathbf{d})=\frac{1}{|\mathbf{d}|} \sum_{n=1}^{|\mathbf{d}| }f\left(d_n\right)~.
\end{equation}

Agents are limited in their capabilities and/or resources, i.e., they might have limited cognitive capacities, limited time, or limited further resources to solve the decision problem. In consequence, they have to collaborate to find a feasible solution to the complex decision problem captured by the task environment. To do so, they decompose the decision problem into $M$ sub-problems $\mathbf{d_m}$, where $m=\{1,\dots,M\}\subset\mathbb{N}$ and $[\mathbf{d}_1,\dots,\mathbf{d}_M]=\mathbf{d}$. For agent $m$, the decisions $\mathbf{d}_m$ represent the area of responsibility (or his or her decision authority), while the complement $\mathbf{d}_{-m} = \mathbf{d} \setminus \mathbf{d}_m$ is referred to as residual decisions. Agents can always observe the solutions to their sub-problem $\mathbf{d}_m$. However, the decisions that are taken by the other agents, i.e.,  solution to the residual decision problem $\mathbf{d}_{-m}$, can only be observed \textit{after} their implementation. 

\begin{figure}
\vspace{0.25cm}
\includegraphics[width=\textwidth]{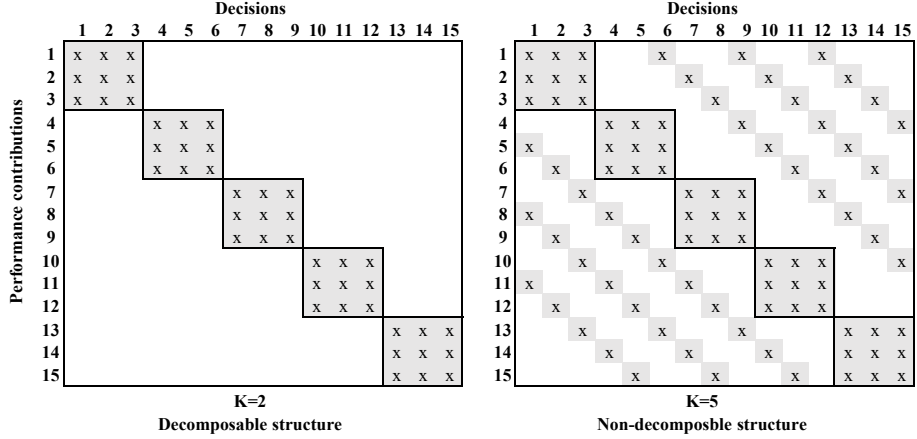}
    \caption{Interdependence matrices}
    \label{fig:matrices}
\end{figure}

This paper considers two stylized interdependence structures presented in Fig. \ref{fig:matrices}, where an \enquote*{x} indicates that a decision and a performance contribution are interdependent. The task allocation indicated by black lines is used as a benchmark scenario that best fulfills the mirroring hypothesis \cite{sanchez1996}. The considered structures are characterized by $K=2$ and $K=5$ that represent a fully decomposable and non-decomposable decision problem, respectively. 
The decomposition of the decision problem can be adapted over time, i.e., agents might trade decisions. In particular, in every period $t\Mod \tau = 0$ agents have the possibility to trade decisions (Sec. \ref{sec:auction}), while in periods $t \Mod \tau \neq 0$ agents seek to maximize their utility given the currently active decomposition of the decision problem (Sec. \ref{sec:decisions}), where $\tau\in \mathbb{N}$. The decomposition in period $t=1$ follows a random process that allocates decisions equally, so that $|\mathbf{d}_m|=N/M$. 
\subsection{Utility Functions and Hillclimbing Mechanism}
\label{sec:decisions}
The performance contributions of agent $m$'s own ($\mathbf{d}_{mt}$) and residual decisions ($\mathbf{d}_{-mt}$) in $t$ are denoted by $P(\mathbf{d}_{mt})$ and $P(\mathbf{d}_{-mt})$, respectively. For the computation of the performances see Eq. \ref{eq:performance}.
The organization employs a linear outcome-based incentive scheme that shapes the agents' utility functions. In particular, the parameters $\alpha\in \mathbb{R^+}$ and $\beta\in\mathbb{R^+}$ are used to weight the agents' own and residual performances, respectively, where $0 \leq \alpha, \beta \leq 1$ and $\alpha+\beta=1$ Agent $m$'s utility at period $t$ is formalized by 
\begin{equation}
    U_{}(\mathbf{d}_{mt},\mathbf{d}_{-mt}) = \alpha \cdot P\left( \mathbf{d}_{mt} \right) + \beta \cdot  P\left( \mathbf{d}_{-mt} \right) ~.
\end{equation}
In every period $t\Mod\tau\neq 0$, the seek to maximize their performance by employing a hillclimbing algorithm \cite{cormen2009}. To do so, agents can either change on random decision in their area of responsibility or stick with the status-quo, if a change in the decisions does not promise to lead to a higher utility. In particular, agent $m$ discovers a solution $\mathbf{d}^{\ast}_{mt}$ to his or her partial decision problem in period $t$, that has a Hamming distance of $1$ to the solution $\mathbf{d}_{mt-1}$, i.e., $\mathbf{d}_{mt}^{\ast}$ is different from $\mathbf{d}_{mt-1}$ in exactly one position (one decision) \cite{Leitner2014}.

Direct communication among agents is omitted in this phase, which is why agent $m$ has no information about the other agents' decision but relies on the other agents' decisions from the previous period, $\mathbf{d}_{-mt-1}$, to compute the utility. Agent $m$ makes his or her decision in period $t$ according to the following rule:
\begin{equation}
\label{eq:decision-rule}
 \mathbf{d}_{mt} := 
    \begin{cases}
    \mathbf{d}_{mt-1}   &   \text{if } U(\mathbf{d}_{mt-1},\mathbf{d}_{-mt-1}) \geq U(\mathbf{d}^{\ast}_{mt},\mathbf{d}_{-mt-1})~, \\
    \mathbf{d}^{\ast}_{mt} & \text{otherwise .}
    \end{cases}
\end{equation}
The solution to the decision problem that is implemented in period $t$ is the concatenation of the individual decisions made by all $M$ agents, 
\begin{equation}
\label{eq:bitstring}
\mathbf{d}_t = \left[ {\mathbf{d}_{1t}}, \dots, \mathbf{d}_{Mt}\right]~,
\end{equation}
\noindent and the performance achieved by the organization in $t$ is $P(\mathbf{d}_t)$ (Eq. \ref{eq:performance}).

\subsection{Learning Mechanism} 
\label{sec:learning}

Agents are aware of interdependencies among decisions, but they do not know the exact structure. However, agents are endowed with beliefs about the interdependencies that can be updated in all periods $t \Mod \tau \neq 0$. We formalize agent $m$'s belief about the interdependencies between decisions $i$ and $j$ in period $t$ by $b_{mt}^{ij} \in \mathbb{R}$, where $i,j = \{1,\dots,N\} \subset \mathbb{N}$, $i\neq j$, and $0\leq b_{mt}^{ij} \leq 1$. The beliefs follow a Beta distribution $B( p_{mt}^{ij}, q_{mt}^{ij})$, where 
\begin{equation}
\label{eq:beliefs}
    b_{mt}^{ij}=\frac{p_{mt}^{ij}}{p_{mt}^{ij}+q_{mt}^{ij}}~.
\end{equation}
For the initial beliefs, $p_{m1}^{ij} = q_{m1}^{ij} = 1$ so that $b_{m1}^{ij} =0.5$. During the runtime, agent $m$'s makes a decision in his or her area of responsibility and fixes the decisions $\mathbf{d}_{mt}$ to be implemented in $t$ (see Sec. \ref{sec:decisions}). If agent $m$ decides to change a decision so that $\mathbf{d}_{mt}:=\mathbf{d}_{mt}^{\ast}$ (Eq. \ref{eq:decision-rule}), beliefs can be updated in the following way:
\begin{enumerate}
    \item Let us denote the decision that has been flipped by agent $m$ in $t$ by $i$, where $d_{it} \in \mathbf{d_{mt}}$.
    \item After the implementation of the decisions $\mathbf{d}_{mt}$, agent $m$ observes the performance contributions of all decisions within his or her area of responsibility.
    \item Whenever agent $m$ observes that the performance contribution of decision $j$ changes from period $t-1$ to period $t$ if the decision $i$ is flipped, $p_{mt}^{ij}$ is updated according to Eq. \ref{eq:update-positive}, otherwise  $q_{mt}^{ij}$ is updated according to Eq. \ref{eq:update-negative}, where $d_{it}, d_{jt} \in \mathbf{d}_{mt}$:  
\end{enumerate}
\begin{subequations}
\begin{eqnarray}
    \label{eq:update-positive}
    p_{mt}^{ij} & = &
    \begin{cases}
      p_{mt-1}^{ij}+1 &   \text{if } f(d_{jt}) \neq f(d_{jt-1})~,\\
      p_{mt-1}^{ij}        &              \text{otherwise .}
    \end{cases} \\
    \label{eq:update-negative}
    q_{mt}^{ij}& = &
    \begin{cases}
      q_{mt-1}^{ij}+1 &   \text{if } f(d_{jt}) = f(d_{jt-1})~,\\
      q_{mt-1}^{ij}        &              \text{otherwise .}
    \end{cases}
\end{eqnarray}
\end{subequations}
\begin{enumerate}
    \setcounter{enumi}{3}   
    \item Agent $m$ recomputes the beliefs according to Eq. \ref{eq:beliefs}.
\end{enumerate}
Please note that agents can only observe the performance contributions \textit{within their own} areas of responsibility. If the decision problem is decomposed so that there are interdependencies with decisions from \textit{outside} an agent's area of responsibility, there might be external influence on performance contributions that the agent cannot identify as such. Therefore, the agents might wrongly induce the existence of interdependencies from their observations.
\subsection{Self-Organization Mechanism}
\label{sec:auction}
In all periods $t \mod \tau = 0$, agents are granted the possibility to re-organize the decomposition of decisions, which we refer to as self-organization. In these periods, agents might follow two different strategies: First, agents might be myopic and trade decisions that increase their utility immediately, referred to as utility-based offers and bids. Second, agents might be rather long-sighted and aim to maximize the interdependencies within their decision problems (internal interdependencies) and minimize the interdependencies with the other agent's decisions (external interdependencies), referred to as interdependence-based offers and bids.  

To account for limitations in resources, every agent is characterized by a maximum capacity $C_{m}$ that indicates the maximum number of decisions that agent $m$ can handle at a time. $C_m$ can be interpreted in terms of maximum cognitive capacity or maximum financial resources, time, manpower, etc., that are available to solve decision problems. The two bidding strategies are described below, whereby agent $m$ offers a decision $i=\{1,\dots,N\} \subset \mathbb{N}$ in and auction and agents $r=\{1,\dots,m-1,m+1,\dots,M\} \subset \mathbb{N}$ submit bids for this decision.
\subsubsection{Computation of utility-based offers and bids.}
Following this strategy, agents aim at maximizing their utility by offering decisions with a relatively low utility to other agents and by biding for decisions with a high utility. The self-organization process is organized as follows:
\begin{enumerate}
    \item Agent $m$ identifies the decision $i$ in his or her own area of responsibility that is associated with the minimum performance contribution and the minimum trading price for this decisions, according to
\end{enumerate}
\begin{equation}
\label{eq:minperf}
   {\rho}_{mt}^{i}=\min_{\forall i: {d}_{it} \in \mathbf{d}_{mt}} \left(f({d}_{it}) \right)~.
\end{equation}
\begin{enumerate}
    \setcounter{enumi}{1}   
    \item Agent $m$ indicates at an auction platform that the decision $i$ is offered in the current auction round. The minimum price for trading this decision is equal to ${\rho}_{mt}^{i}$ (Eq. \ref{eq:minperf}). The available decision is referred to as offer.
    \item Once all $M$ agents have submitted their offers, agents start placing their bids.\footnote{Agent $m$ cannot bid for his or her own offer of decision $i$, but only for the other decisions that are offered in the current auction round, and agents cannot observe the other agents' bids.} However, agents can only participate in the auction if they have free resources, i.e., agent $r$ proceeds with the next step iff $|\mathbf{d}_{rt}| < C_r$.
    \item Agents $r$ submits the expected performance contribution of decision $i$ as a bid in period $t$. Since the offered decision is outside of agent $r$'s area of responsibility, she or he can only estimate the related performance contribution according to 
    \begin{equation}
        \label{eq:perf-basedbids}
        \bar{\rho}_{rt}^{i}= f(d_{it}) + \epsilon~,
    \end{equation}
    where $\epsilon \sim N(0,\sigma)$ represents agent $r$'s error in the estimation of decision $i$'s performance contribution.
\end{enumerate}
%
\subsubsection{Computation of interdependence-based offers and bids.}
Following this strategy, agents do not aim at immediately maximizing their utility but they aim at maximizing the interdependencies within their own areas of responsibility. The self organization is organized as a sequential process:
\begin{enumerate}
    \item Agent $m$ identifies the decision $i$ in his or her own area of responsibility that is associated with the minimum belief for internal interdependencies. The minimum trading price for this decision is computed according to Eq. \ref{eq:mink}.\footnote{For simplicity, the average belief about interdependencies between decision $i$ and decisions $j$ in agent $m$'s area of responsibility is used as the minimum price.}
\end{enumerate}
\begin{equation}
\label{eq:mink}
   {\rho}_{mt}^{i} = \min_{\forall i: d_{it} \in \mathbf{d}_{mt}}\left( \frac{1}{|\mathbf{d}_{mt}|-1} \sum_{\substack{\forall j: d_{jt} \in \mathbf{d_{mt}} \\ j \neq i}} b_{mt}^{ij} \right)
\end{equation}
\begin{enumerate}
    \setcounter{enumi}{1}   
    \item Agent $m$ indicates at an auction platform that the decision $i$ that fulfils Eq. \ref{eq:mink} is offered in the current auction, and that the minimum price for trading this decision is equal to ${\rho}_{mt}^{i}$. The available decision is referred to as offer.
    \item Agent $r$ proceeds with step 4 and places bids iff $|\mathbf{d}_{rt}| < C_r$.
    \item Agents $r$ submit the average belief about the interdependencies between the offered decision $i$ with the decisions within his or her area of responsibility $\mathbf{d}_{rt}$ as a bid in period $t$. Agent $r$'s bid for decision $i$ in $t$ is formalized by
    \begin{equation}
        \label{eq:k-basedbids}
        \bar{\rho}_{rt}^{i}= \frac{1}{|\mathbf{d}_{rt}|} \sum_{{\forall j: d_{jt} \in \mathbf{d_{rt}}}} b_{rt}^{ij}
    \end{equation}
\end{enumerate}
\subsubsection{Task re-allocation.}
Once all agents submitted their bids, for every offer $i$ there are at most $M-1$ bids. Recall, agent $m$ offered decision $i$ at a minimum price of ${\rho}^{i}_{mt}$ (according to either Eq. \ref{eq:minperf} or \ref{eq:mink}) and the other agents submitted bids $\bar{\rho}^{i}_{rt}$ (according to either Eq. \ref{eq:perf-basedbids} or \ref{eq:k-basedbids}). 
Let us denote the set of bids for decision $i$ in period $t$ by $P_t^i$, the maximum bid for decision $i$ in period $t$ by $\bar{\rho}^{i}_{r^{\ast}t}= \max_{\bar{\rho}^{i}_{rt}\in P_t^i} (\bar{\rho}^{i}_{rt})$, and the agent placing this bid by $r^{\ast}$. The decisions are (re-)allocated as follows:
\begin{enumerate}
    \item If the the maximum bid $\bar{\rho}^{i}_{r^{\ast}t}$ is equal to or exceeds the minimum price ${\rho}^{i}_{mt}$, the decision $i$ is re-allocated from agent $m$ to agent $r^{\ast}$ according to
    \begin{subequations}
    \begin{eqnarray}
      \mathbf{d}_{mt} & := & \mathbf{d}_{mt-1} \setminus \{ d_{it-1} \} ~\text{and}\\
        \mathbf{d}_{r^{\ast}t} & := & \left[ {\mathbf{d}_{r^{\ast}t-1}},d_{it-1}  \right]~,
        \end{eqnarray}
    \end{subequations}
    \noindent where $\setminus$ indicates the complement. If the second highest bid exceeds the minimum price, agent $r^{\ast}$ gets charged (a function of) the second highest bid, otherwise agent $r^{\ast}$ gets charged (a function of) the minimum price. 
    \item If the the maximum bid $\bar{\rho}^{i}_{r^{\ast}t}$ does not exceed the minimum price ${\rho}^{i}_{mt}$, agent $m$ remains responsible for the decision $i$, so that 
    \begin{equation}
        \mathbf{d}_{mt}:=\mathbf{d}_{mt-1}~.
    \end{equation}
\end{enumerate}

\subsection{Scenarios and Performance Measure}

This paper puts particular emphasis on the effects of variations in three parameters and considers
\begin{itemize}
\item two different levels/structures of complexity, $K=2$ and $K=5$ for decomposable and non-decomposable structures, respectively (Sec. \ref{sec:task-environment}),
\item three different incentive parameters, where $\{\alpha,\beta\}$ is set to $\{1,0\}$, $\{0.5,0.5\}$, and $\{0.25,0.75\}$ for individualistic, balanced, and altruistic incentive systems, respectively (Sec. \ref{sec:decisions}), and 
\item two strategies for self-organization, i.e. performance- and interdependence-based offers and bids (Sec. \ref{sec:auction}).
\end{itemize}
The number of decisions and the number of agents are set to $N=15$ and $M=5$. Agents are granted the possibility to self-organize every $\tau=25$ periods, whereby every agent's cognitive capacity is limited so that at most $C_m=5$ decisions can be handled at one point in time. In the case of performance-based offers and bids, the standard deviation of errors that agents make when predicting expected performances is set to $\sigma=0.05$ (Eq. \ref{eq:perf-basedbids}). In addition to scenarios that consider self-organization (Sec. \ref{sec:auction}), a benchmark scenario is included in the results, in which tasks are allocated to agents in line with the mirroring hypothesis, i.e., according to the technical characteristics of the tasks so that cross-dependencies are minimized. The task allocation for the benchmark scenario is indicated by bold lines in Fig. \ref{fig:matrices}. For all scenarios, the achieved performances are observed for $T=500$ periods, and for each scenario, we run $S=800$ repetitions.  

The performance achieved in simulation round $s=\{1,\dots,S\}\subset \mathbb{N}$ and period $t$ is denoted by $P(\mathbf{d}_{ts})$ (Eq. \ref{eq:performance}). To assure comparability across simulation runs, the observed performance $P(\mathbf{d}_{ts})$ is normalized by the maximum attainable performance in that scenario, $P(\mathbf{d}^{\ast}_s)$. Finally, the normalized mean performance achieved in periods $t$ is reported as performance measure:
\begin{equation}
\label{eq:performancemeasure}
\tilde{P}_t=\frac{1}{S} \sum_{s=1}^{S} \frac{P(\mathbf{d}_{ts})}{P(\mathbf{d}^{\ast}_{s})}~.
\end{equation}
\section{Results and Discussion}
\label{sec:results}
\begin{figure}
\includegraphics[width=\textwidth]{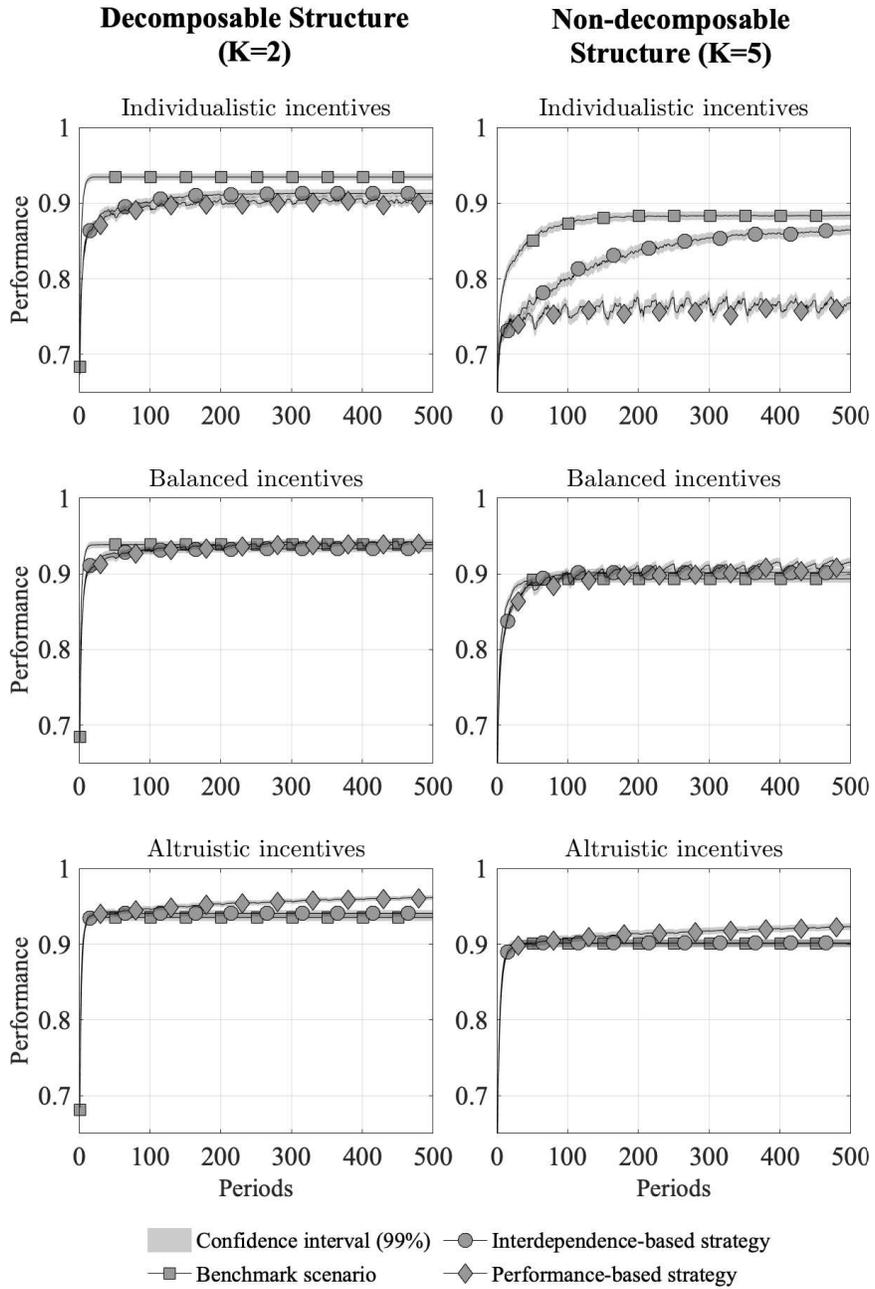}
    \caption{Organizational performance for different incentives schemes and different self-organization strategies}
    \label{fig:results}
\end{figure}

This paper particularly focuses on the analysis of the normalized performance achieved in organizations that follow an evolutionary approach to organizational design and allows departments to self-organize, i.e., to autonomously allocate tasks. Figure \ref{fig:results} presents performance achieved over time (Eq. \ref{eq:performancemeasure}) for scenarios in which either individualistic, balanced or altruistic incentive schemes are effective. The strategies of utility-based and interdependence-based offers and bids are indicated by diamonds ($\MyDiamond$) and circles ($\tikzcircle[fill=gray]{3pt} $), respectively. In addition, the results include the performance achieved in the benchmark scenario ($\greysquare$). Recall that in the benchmark scenario, the allocation of tasks is fixed throughout the simulation runs, so that internal interdependencies are maximized (according to the solid lines in Fig. \ref{fig:matrices}). The shaded areas around the curves indicate confidence intervals at the $99\%$-level.

\subsubsection{Individualistic Incentive Schemes.} 
The results indicate that for \textit{decomposable structures} and individualistic incentive schemes, it appears to be beneficial to implement the benchmark scenario immediately since the performance almost immediately increases to its limit. Evolutionary approaches increase at a slower speed and have a lower limit, whereby no significant differences between the two self-organization strategies can be observed. For \textit{non-decomposable structures}, the results similarly indicate that the highest performance can be achieved in the benchmark scenario. The results for the evolutionary approaches to organizational design are pretty much in line with what is expected: First, for individualistic incentive systems, the results prove that it is beneficial for performance if task allocation follows the task's technical characteristics \cite{sanchez1996}. Second, the results show that for myopic agents who follow a utility-based strategy \enquote*{sacrifice the long-run to the short-run} \cite{levinthal1993,ridge2014}, so that the performance finally achieved is significantly below the performances achieved in the other scenarios. For cases in which agents aim at optimizing the task allocation and thereby make changes to task allocation that do not pay off immediately but might be beneficial in the long run, a significantly higher performance can be achieved \cite{laverty1996}. However, time-preferences only appear to affect the performance in the case of non-decomposable tasks, whereas in the decomposable case, myopia appears not to affect performance.

\subsubsection{Balanced Incentive Schemes.}
For both \textit{decomposable} and \textit{non-decomposable tasks}, switching from individualistic to balanced incentive schemes appear to affect the speed of performance improvement for both evolutionary approaches to organizational design. In particular, it can be observed that the performance increases to the level of the benchmark scenario. This observation is supported by previous research on contracting externalities which finds incentive systems that account for interdependencies have a chance to increase performance \cite{gibbons1998}. The results also indicate that switching from individualistic to balanced incentive schemes renders the effect of time-preferences trivial, since for both utility-based (myopic) and interdependence-based (long-term oriented) strategies in self-organization, almost identical performances can be achieved, whereby decision-makers tend to respond stronger to changes in the incentive scheme if the task is non-decomposable. This finding is robust across all analyzed interdependence structures. A large proportion of the literature on incentive schemes is concerned with designing specific incentive mechanisms that make sure that managers have strong incentives to make present value-maximizing decisions (e.g., by rules of accrual accounting) \cite{rajan2009,pfeiffer2009,dutta2005}. The results presented here show that linear incentive schemes can provide these incentives in the context of autonomous evolutionary design approaches.

\subsubsection{Altruistic Incentive Schemes.}
In the case of altruistic incentive schemes, it can be observed that the performances achieved in the benchmark scenario and the case of the interdependence-based self-organization strategy are at the same level throughout the observation period. However, the results indicate that the utility-based approach in autonomous evolutionary design appears to pay off in the long run since the achieved performance is even higher than the benchmark performance. This is an important finding since the results show that linear incentive schemes can mitigate potential adverse effects caused by myopic decision-makers like it was observed for balanced incentive schemes. The results further indicate that switching from balanced to altruistic incentive schemes, seemingly shortsighted decisions in autonomous evolutionary approaches to organizational design, are desirable not only in short but also in the long run. This means that the problems of how to optimally integrate the managers' long-term and short-term thinking \cite{hrebiniak1986} and how to provide incentives for long-term orientation \cite{narayanan1985} might be solved by linear incentives.

\section{Conclusion}
\label{sec:conclusion}

In this paper, a model of evolutionary organizational design is implemented that includes \textit{autonomous} task allocation at the micro-level. In particular, when allocating tasks, agents have different time-preferences, i.e., they follow either a utility-based or an interdependence-based strategy. If agents follow the former approach, they aim at increasing their utility immediately by trading decisions. In contrast, in the latter case, agents optimize for the technical characteristics of the task so that their utility increases in the long run. Given this organizational setup, the paper mainly analyzes the achieved performance if either individualistic, balanced, and altruistic incentive schemes are effective. The results provide some insightful extensions to previous research on this issue. First, considering organizational performance, altruistic incentive schemes are to be preferred over individualistic incentive schemes. This is particularly pronounced for non-decomposable tasks. Second, implementing balanced or altruistic incentive schemes and avoiding individualistic incentives might mitigate the adverse effects caused by myopic decisions. Third, irrespective of the technical characteristics of the task, altruistic incentive schemes appear substantially interact with the self-organization mechanisms so that, from the organization's perspective, myopic decisions are favorable in the long run.

This research is the first step towards a theory of (behavioral control in) autonomous evolutionary organizational design. Further research should analyze the effects of task structures and additional self-organization strategies on the performance and explore the resulting performance and take into account the emergent organizational structure.

%
%
%
\bibliography{bib}
\bibliographystyle{splncs03}

\end{document}